\documentclass{article}

\usepackage{arxiv}

\usepackage[utf8]{inputenc} 
\usepackage[T1]{fontenc}    
\usepackage{hyperref}       
\usepackage{url}            
\usepackage{booktabs}       
\usepackage{amsfonts}       
\usepackage{nicefrac}       
\usepackage{microtype}      
\usepackage{lipsum}		
\usepackage{physics}
\usepackage{comment}
\usepackage[toc,page]{appendix}
\newcommand{\nl}{\nonumber \\}
\title{$1/\epsilon$ problem in resurgence}

\author{
  Naohisa Sueishi\\
  Department of Physics\\
  Nagoya University\\
  \texttt{sueishi@eken.phys.nagoya-u.ac.jp} \\
}

\begin{document}
\maketitle

\begin{abstract}
This paper considers the $1/\epsilon$ problem, which is the divergent behavior of the ground state energy of asymmetric potential in quantum mechanics, which is calculated with semi-classical expansion and resurgence technique. Using resolvent method, It is shown that including not only one complex bion but multi-complex bion and multi-bounce contributions solves this problem. This result indicates the importance of summing all possible saddle points contribution and also the relationship between exact WKB and path integral formalism.

\end{abstract}


\section{Introduction}
For many quantum theories, perturbative expansion is used to compute physical quantities. Nevertheless, generally the radius of convergence of such the series is zero -- it means the series is not convergent but asymptotic. This result was first suggested by Dyson\cite{PhysRev.85.631} in QED and it was proved that this is due to the fact that the number of Feynman diagrams grows factorially as the order of the perturbative series grows\footnote{If a perturbative series is convergent, it indicates there are some symmetries which cancel a large number of diagrams, e.g. supersymmetry.}\cite{doi:10.1098/rspa.1952.0149}\cite{PhysRevLett.37.117}\cite{KumarPatnaik1990}. Resurgence theory gives a systematic treatment to solve this problem and also reveals hidden relationship between perturbative and nonperturbative effects.

In quantum theory, 
perturbative expansion around classical solutions take the following type of series:
\begin{align}
  Z(\hbar) &=\int_{periodic}\mathcal{D}\phi\;e^{-\frac{S[\phi]}{\hbar}}\\
  &= \sum_n a_n\hbar^n + e^{-\frac{S_b}{\hbar}}\sum_n b_n\hbar^n + e^{-\frac{2S_b}{\hbar}}\sum_n c_n\hbar^n+...
\end{align}
$S_b$ is the action of a nonperturbative saddle which satisfies periodic boundary condition for imaginary time, like bion(instanton and anti-instanton) configuration. This type of series is called trans-series and each of the series is asymptotic. A method to make sense of factorially divergent series is the Borel transform, which makes the series convergent. The Laplace transform of the Borel transform is called Borel sum, which has the same asymptotic expansion as the original series but is the convergent function: that is, for a series expansion such as
\begin{align}
  Z(\hbar)=e^{-\frac{A}{\hbar}}\sum_{n=0}^{\infty}a_n\hbar^{n+\alpha}\;\;\;\;\alpha\notin \{-1,-2,-3,...\}
\end{align}
the Borel transform of this series is defined by
\begin{align}
  B[Z](z) \equiv \sum_{n=0}^{\infty}\frac{a_n}{\Gamma(n+\alpha)}(z-A)^{n+\alpha-1}
\end{align}
The Borel summation:$\mathcal{S}[Z]$ is defined as
\begin{align}
  \mathcal{S}[Z](\hbar)\equiv\int_A^{\infty e^{i\theta}} e^{-\frac{z}{\hbar}}B[Z](z)\dd{z}\;\;\;\;\theta=arg(\hbar)
  \label{Borel2}
\end{align}
Borel summation often has Borel ambiguity when poles exist on the integral path of the Laplace transform. The resurgence theory claims that these ambiguities arisen from the each sector of trans-series are cancelled\footnote{This claim is coming from the equivalence of Borel summation and integration on Lefschetz thimble.}. From this cancellation, we can obtain the information of other saddle points, which mean the nonperturbative effects. \cite{Aniceto2015a}\cite{Serone2017a}\cite{Delabaere1997}\cite{Dunne2014a} The resurgence method can be applied to various systems, which include QFT\cite{Honda2016}\cite{Klaczynski2016}\cite{Cherman2015} and string theory\cite{Aniceto2012}\cite{Codesido2019} because perturbative expansion is quite general method in physics.
Further, the relationship between resurgence theory and Picard-Lefschetz theory\cite{Yuya} tells us a new perspective in path integral formalism. In particular, the role of complexified classical solutions in evaluating the path integral. \cite{Behtash}\cite{Behtash2015}\cite{1510.00978}\cite{Dunne2015}\cite{Serone2017}\cite{Fujimori2016} This is precisely related to the work of Witten on Chern-Simons theory \cite{Witten2011}, which claims the complexification of the phase space formalism of path integral.

The resurgence technique has been enough to allow one to uncover the hidden structures in quantum theory. However, there remains unsolved problems, one of which is the $1/\epsilon$ problem. The papers \cite{1510.00978} \cite{1510.03435} discuss supersymmetric quantum mechanics and show Borel ambiguity corresponding to the perturbative expansion around the vacuum is cancelled by a complex bion\footnote{This is a non-BPS periodic solution satisfying complexified Newton's equation.}. It explains the nonperturbative shift of the ground state energy (dynamical SUSY breaking).
But the expression of the ground state energy in \cite{1510.03435} becomes singular when the Lagragian is deformed to purely bosonic case. This is called $1/\epsilon$ problem.

In this paper, we show a method to solve the problem and the reason behind the prescription. It indicates the relationship between path integral formalism and exact WKB\cite{Dorigoni2019}
\cite{Costin2008}
\cite{Iwaki2014}, which is a resurgence method to analyze the structure of differential equation\footnote{Originally resurgence by Ecalle was used for Stokes phenomena of differential equations.}.

This paper is organized as follows: The rest of this section explains SUSY QM briefly and what $1/\epsilon$ problem is. Sections 2 and 3 are dedicated to the detailed calculation of the partition function and the leading noperturbative contribution of the ground state energy.
In section 4 discusses the relation between Fredholm determinant and exact WKB calculous. In section 5 gives conclusions and summary.
\subsection{$1/\epsilon$ problem}
Consider the Witten model:
\begin{align}
  S&=\int_{-\infty}^{\infty}\dd{t}\qty(\frac{1}{2}\dot{x}^2-\frac{1}{2}(W'(x))^2+i\eta^\dagger \dot{\eta}+W''(x)\eta^\dagger\eta)
\end{align}
where $W(x)$ is a superpotential. The Hamiltonian of this system can be written in terms of only bosonic variable after projecting to fermion number eigenstates:
\begin{align}
  H &= \mqty(H_+ & 0 \\ 0 & H_-)\\
  H_\pm &= \frac{1}{2}p^2+V_\pm(x)\\
  &=\frac{1}{2}p^2+\frac{1}{2}(W'(x))^2\mp\frac{1}{2}\hbar W''(x)\\
\end{align}
The term $\mp\frac{1}{2}\hbar W''$ comes from the fermion terms. The Euclidean path integral is $Z=\int \mathcal{D}x\;e^{-\frac{S_E}{\hbar}}$ in this notation.
Now, set the superpotential to $W(x) = \frac{1}{3}x^3-a^2x$. It gives
\begin{align}
  H_\pm=\frac{1}{2}p^2+\frac{1}{2}(x^2-a^2)^2\mp \hbar x
\end{align}
The zero energy eigenstate is
\begin{align}
  \bra{x}\ket{0}&=e^{-\frac{W(x)}{\hbar}}=e^{-\frac{1}{\hbar}\qty(\frac{1}{3}x^3-a^2x)}
\end{align}
This state is not normalizable in the real axis: $(-\infty,\infty)$. Therefore supersymmetry of this system is dynamically broken. the SUSY breaking is due to nonperturbative effects, because the perturbative contribution to the ground state energy vanishes. To examine the resurgence structure, we need to introduce deforming parameter $\epsilon$ here:
\begin{align}
  H_\pm=\frac{1}{2}p^2+\frac{1}{2}(x^2-a^2)^2\mp\epsilon \hbar x
\end{align}
When $\epsilon=1$, this system returns to the original SUSY Hamiltonian. 

The perturbative expansion of ground state energy\footnote{Of course when we set $\epsilon=1$, the all coefficients are vanished by supersymmetry.} reads \cite{Behtash2017}
\begin{align}
E_{0,pert}&=\sum_{n=0}^\infty a_n\hbar^n\\
a_n&=-\frac{6^{-\epsilon+1}}{2\pi}\frac{\Gamma(n-\epsilon+2)}{\Gamma(1-\epsilon)}\frac{1}{(2S_I)^n},
\end{align}
where $S_I=\frac{3\hbar}{4a^3}$ is the one instanton action. 
The Borel summation of this series is
\begin{align}
  -\frac{6^{-\epsilon+1}}{2\pi}\frac{1}{\Gamma(1-\epsilon)}\int_0^\infty e^{-z}\frac{z^{-\epsilon+1}}{1-\frac{z}{2S_I}}dz
\end{align}
It has this Borel ambiguity, 
\begin{align}
  \Im \mathcal{S}[E_{0,pert}](\hbar)=\mp \frac{1}{2}6^{-\epsilon+1}\frac{1}{\Gamma(1-\epsilon)}2S_I^{-\epsilon}e^{-2S_I}
  \label{SUSYBorelambiguity}
\end{align}
$\pm$ is corresponded to the sign of $\Im(\hbar)$. 

Resurgence theory claims the Borel smbiguity from a perturbative expansion is cancelled by considering the contribution of other saddle points. In this system, there are classical solutions called the \textit{complex bion} \cite{1510.03435}:
\begin{align}
  x_{cb}(t)=x_1-\frac{x_1-x_T}{2}\coth\qty(\frac{\omega_{cb}t_0}{2})\qty(\tanh\qty(\omega_{cb}\frac{t+t_0}{2})-\tanh\qty(\omega_{cb}\frac{t-t_0}{2})),
\end{align}
where $x_1$ is the position of potential minimum, $x_T=-x_1+i\sqrt{\frac{\epsilon \hbar}{-x_1}}$ is the complex turning point, $\omega_{cb}=\sqrt{V''(x_1)}$, $t_0=\frac{2}{\omega_{cb}}\mathrm{arccosh}\qty(\sqrt{\frac{3}{1-V''(x_T)/\omega_{cb}^2}})$.\\

If we flip the parameter $\epsilon$ to $-\epsilon$, the complex bion becomes bounce, which is the other classical solution. However, naively the bounce is a high energy configuration and therefore is not expected to contribute to the ground state energy.
In \cite{1510.03435}, the nonperturbative effect for the ground state energy is calculated by:
\begin{align}
  E_0=-\frac{1}{\beta}\log Z&=-\frac{1}{\beta}\log(Z_0+Z_{cb}+Z_{2cb}+...)\\
  &\simeq-\frac{1}{\beta}\log Z_0-\frac{1}{\beta}\qty(\frac{Z_{cb}}{Z_0})
\end{align}
$Z_0$ and $Z_{cb}$ are partition functions corresponding to the vauum and one complex bion solution, respectively.
It suggests one complex bion is enough to obtain the leading nonperturbative contribution for ground state energy. Using this method, the nonperturbative contribution, which is from one complex bion is 
\begin{align}
  \label{Icb}
  I_{cb}=-\frac{1}{\beta}\frac{Z_{cb}}{Z_0}=\frac{1}{2\pi}\qty(\frac{\hbar}{16a^3})^{\epsilon-1}\qty(-\cos(\epsilon\pi)\Gamma(\epsilon)\pm i\frac{\pi}{\Gamma(1-\epsilon)})e^{-2S_I}
\end{align}
Again, $\pm$ is corresponded to the sign of $\Im(\hbar)$.\footnote{This ambiguity is not Borel ambiguity but coming from Stokes phenomena of quasi-moduli integral. However, It does not conflict with the resurgence claim because of the equivalence of Borel summation and integral on the Lefschetz thimble.} Combining the two results shows
\begin{align}
  \Im(S[E_{0,pert}]+I_{cb})=0
\end{align}
Therefore the ground state energy is
\begin{align}
  \label{CBE02}
  E_0&=-\frac{1}{2\pi}\qty(\frac{-\hbar}{16a^3})^{\epsilon-1}\Gamma(\epsilon)e^{-\frac{8a^3}{3\hbar}}\cos(\epsilon\pi)\\
  &=-\frac{1}{2\pi}\qty(\frac{\hbar}{16a^3})^{-1}\qty(\frac{1}{\epsilon}-\gamma+\order{\epsilon})e^{-2S_I}
  \label{CBE0}
\end{align}
The Borel ambiguity from perturbative expansion is exactly cancelled by the other ambiguity from nonperturbative (complex bion) saddle. However, this expression has two strange facts: (i). This is singular as $\epsilon\rightarrow 0$, which is the case of symmetric double well potential. (ii). In the case of symmetric double well, the nonperturbative contribution of the ground state energy is coming from one-instanton.\cite{Coleman2011} So even if we remove the singularity by hand, the result is still incorrect. \\
This is the $1/\epsilon$ problem in deformed SUSY quantum system. The problem does not only occur in the case of tilted double well, but is known to occur in $CP^N$ and sine-Gordon model. \cite{Fujimori2016}\cite{Fujimori2017}\cite{Fujimori2017a}

\section{Our calculation and result}

\subsection{The prescription}
The method to calculate the ground state energy is based on Euclidean path integral with periodic boundary condition.
\begin{align}
  Z(\beta)&= \int_{\textrm{periodic},\beta} \mathcal{D}x e^{-\frac{S_E[x]}{\hbar}}
\end{align}
For computing the partition function $Z(\beta)$ properly, all the classical solutions whose period are $\beta$ should be taken into account. Also the partition function should be invariant under $\epsilon\rightarrow -\epsilon$ because the spectrum doesn't change by this reflection. Therefore we have to \textit{multi-complex bion and multi-bonce} with finite $\beta$ to obtain the correct contribution.
The calculation in \cite{Behtash2017}, they considered only one complex bion and $\beta\rightarrow\infty$ limit first. These procedures do not treat the symmetry properly and lead $1/\epsilon$ problem as we show in the next section.

\subsection{The resolvent method}
The ground state energy is worked out by employing resolvent method\cite{Zinn-Justin2004}
\begin{align}
  \int_0^\infty Z(\beta)e^{\beta E}\dd{\beta}&=\int_0^\infty \sum_n e^{-\beta(E_n-E)}\dd{\beta}\\
  &=\sum_n \frac{1}{E_n-E}\\
  &=\tr\frac{1}{H-E}=G(E)
\end{align}
The trace of resolvent $G(E)$ can be expressed as 
\begin{align}
  -\pdv{E}\log D=G(E),
\end{align}
where $D(E)=\det(H-E)$ is the Fredholm determinant. The poles of $G(E)$ and the zeros of $D(E)$ encode the spectrum of $H$.

\subsection{The partition function}
The partition function is shown to take the form:
\begin{align}
  \label{ZZ0}
  \frac{Z}{Z_0}&=1e^{\beta a \epsilon}+1e^{-\beta a\epsilon}+\sum_{n=1}^\infty\qty(e^{-2S_{I}}\frac{S_{I}}{2\pi}\qty(\frac{\det M_{I}}{\det M_0})^{-1})^n\beta QMI^{n}(\epsilon)\nonumber\\
  &+\sum_{n=1}^\infty\qty(e^{-2S_{I}}\frac{S_{I}}{2\pi}\qty(\frac{\det M_{I}}{\det M_0})^{-1})^n\beta QMI^{n}(-\epsilon)
\end{align}
The first $1e^{\beta a \epsilon}+1e^{-\beta a\epsilon}$ are from stationary classical solutions (vacuum and false vacuum). The factors $e^{\pm\beta a\epsilon}$ come from $e^{-S_{vac}/\hbar}=e^{-V(x_{vac})/\hbar}=e^{\beta a \epsilon}$. The latter terms come from the nonperturbative contributions, which are complex bions and bounces, respectively.
The linear factor $\beta$ is from the translation symmetry of (imaginary) time dependent solutions. $B=e^{-2S_{I}}\frac{S_{I}}{2\pi}\qty(\frac{\det M_{I}}{\det M_0})^{-1}$ is the square of one instanton contribution: bion contribution.
\begin{align}
  &x_I(\tau)=a\tanh a(\tau-\tau_c)\\
  &S_I=\frac{S[x_I,\epsilon=0]}{\hbar}=\frac{4a^3}{3\hbar}\\
  &\frac{\det M_{I}}{\det M_0}=\frac{1}{12}
\end{align}
The exact classical solution is not this instanton but complex bions and bounces (These solutions are interchanged by $\epsilon\rightarrow -\epsilon$
). To calculate the contribution from these solutions for path integral, we have to consider quasi-moduli integral (QMI), which comes from a nearly flat direction in the configuration space.

QMI, comes from the nearly flat direction in complex bion solution, i.e. the separation between instanton and anti-instanton in a complex bion is
\begin{align}
  \tau=2t_0\simeq \frac{1}{2a}\qty(\log\qty(\frac{16a^3}{\epsilon\hbar}\pm i\pi))
\end{align} 
This can be infinite under $\hbar\rightarrow 0$, which leads to  quasi-zero mode\footnote{The existence of this direction comes from the $\hbar$ dependence of our potential.}. Therefore we have to consider the interaction potential $\mathcal{V}$ whose variable is $\tau$. The complex bion itself is understood as the saddle point of $\mathcal{V}$.

The form of quasi-moduli integral for $n$-complex bions is\footnote{There are two quasi-moduli integrals for one complex bion because we consider finite $\beta$ now.}
\begin{align}
  QMI^n(\epsilon)=e^{\beta a \epsilon}&\frac{1}{2n}\qty(\prod_{i=1}^{2n}\int_0^\infty d\tau_i e^{-\mathcal{V}_i(\tau_i)})\delta\qty(\sum_{k=1}^{2n} \tau_k-\beta)\nl
  \mathcal{V}_i(\tau)&=\begin{cases}
    -\frac{16a^3}{\hbar}e^{-2a\tau}+2a\epsilon\tau & (i=odd)\\
    -\frac{16a^3}{\hbar}e^{-2a\tau} & (i=even)
  \end{cases}
  \label{Vbefore}
\end{align}

The factor $\frac{1}{2n}$ in front of the integral arises because the configuration is invariant under cyclic permutation of the $\tau_i$. The factor $e^{\beta a \epsilon}$ comes from changing the off-set because the interaction potential $\mathcal{V}_i$ is determined from the true vacuum (the minimum point) but the two vacua have the potential $e^{\beta a \epsilon}$. Actually this procedure is equivalent to setting the $\mathcal{V}_i(\tau)$ as
\begin{align}
  \mathcal{V}_i(\tau)&=\begin{cases}
    -\frac{16a^3}{\hbar}e^{-2a\tau}+a\epsilon\tau & (i=odd)\\
    -\frac{16a^3}{\hbar}e^{-2a\tau}-a\epsilon\tau & (i=even)
  \end{cases}
\end{align}
and omitting the factor $e^{\beta a \epsilon}$. 
When $i=even$, this integral is ill-defined, but if we do the analytic continuation after performing the integral, it coincides with (\ref{Vbefore}). Hence, if we consider an infinite number of complex bions and bounces instead of one, we can say that the contributions of them enter on the same order of magnitude.

If we set $\mathcal{V}_i=0$ and $\epsilon=0$, this integral becomes
\begin{align}
  \qty(\prod_{i=1}^{2n}\int_0^\infty d\tau_i) \delta\qty(\sum_{k=1}^{2n} \tau_k-\beta)=\frac{1}{(2n-1)!}\beta^{2n-1}
\end{align}
Therefore the partition function is evaluated as
\begin{align}
  \frac{Z}{Z_0}=2\sum_{n=0}^{\infty}\frac{B^n\beta^{2n}}{(2n)!}=2\cosh(\sqrt{B}\beta)
  \label{DIGbion}
\end{align}
This is the result that are obtained by employing the standard dilute instanton gas approximation of symmetric double well potential.\footnote{ we can see the importance of including multi-bion configurations in this approximation. See Appendix \ref{DIG}}

From here we set $a=\frac{1}{2}$ for simplicity. This integral can be evaluated as
\begin{align}
  QMI^n(\epsilon)&= e^{\beta\frac{\epsilon}{2}}\frac{1}{2n}\prod_{i=1}^{2n}\qty(\int_0^\infty d\tau_i e^{-\mathcal{V}_i(\tau_i)})\delta\qty(\sum_{k=1}^{2n} \tau_k-\beta)\\
  &=e^{\beta\frac{\epsilon}{2}}\frac{1}{2n}\prod_{i=1}^{2n}\qty(\int_0^\infty d\tau_i e^{-\mathcal{V}_i(\tau_i)})\frac{1}{2\pi}\int_{-\infty}^\infty dl e^{il\sum_{k=1}^{2n} (\tau_k-\beta)}\\
  &=e^{\beta\frac{\epsilon}{2}}\frac{1}{2n}\frac{1}{2\pi}\int_{-\infty}^{\infty}dl e^{-il\beta}\qty(\qty(\int_0^\infty d\tau e^{(il\tau+\frac{2}{\hbar}e^{-\tau})})\qty(\int_0^\infty d\tau e^{(il\tau-\epsilon\tau+\frac{2}{\hbar}e^{-\tau})}))^n\\
  &=e^{\beta\frac{\epsilon}{2}}\frac{1}{2n}\frac{1}{2\pi i}\int_{-i\infty}^{i\infty}ds e^{-s\beta}\qty(\qty(\int_0^\infty d\tau e^{(s-\epsilon)\tau+\frac{2}{\hbar}e^{-\tau}})\qty(\int_0^\infty d\tau e^{s\tau+\frac{2}{\hbar}e^{-\tau}}))^n
\end{align}
We can show the integral is evaluated as
\begin{align}
  \int_0^\infty d\tau e^{(s-\epsilon)\tau+\frac{2}{\hbar}e^{-\tau}}=e^{\pm i\pi (\epsilon-s)}\qty(\frac{\hbar}{2})^{\epsilon-s}\Gamma(\epsilon-s)
\end{align}
Here the $\pm$ is corresponded to the sign of $\Im(\hbar)$, which comes from Stokes phenomenon of this quasi-moduli integral (See Appendix F in \cite{Fujimori2017a}).

Therefore, the form of quasi-moduli integral for finite $\beta$ becomes
\begin{align}
  QMI^n(\epsilon)=\frac{1}{4\pi i n}e^{\beta\frac{\epsilon}{2}}\int_{-i\infty}^{i\infty}dse^{-s\beta}\qty(\qty(e^{\pm i\pi (\epsilon-s)}\qty(\frac{\hbar}{2})^{\epsilon-s}\Gamma(\epsilon-s))\qty(e^{\pm i\pi (-s)}\qty(\frac{\hbar}{2})^{-s}\Gamma(-s)))^n
\end{align}

\subsection{Calculating the resolvent}
Using $Z_0=\sum_{k=0}^{\infty}e^{-\beta(k+1/2)}\sim e^{-\beta\frac{1}{2}}$(for large $\beta$), (\ref{ZZ0}) can be written as
\begin{align}
  \label{Form,Z}
  Z&=\qty\Big{Z_0e^{\beta\frac{\epsilon}{2}}+\sum_{n=1}^\infty B^n e^{-\beta(1/2)}\beta e^{\beta\frac{\epsilon}{2}}\frac{1}{4\pi in}\int_{-i\infty}^{i\infty}ds e^{-s\beta}(I(s,\epsilon)I(s,0))^n}+\qty\Big{(\epsilon\rightarrow-\epsilon)}\\
  B&=e^{-2S_{I}}\frac{S_{I}}{2\pi}\qty(\frac{\det M_{I}}{\det M_0})^{-1}=\frac{e^{-\frac{1}{3\hbar}}}{2\pi}\frac{2}{\hbar}\\
  I(s,\epsilon)&=e^{\pm i\pi (\epsilon-s)}\qty(\frac{\hbar}{2})^{\epsilon-s}\Gamma(\epsilon-s)
  \label{Form:I}
\end{align} 
The Laplace transform gives the trace of resolvent $G(E)$.
\begin{align}
  G(E)&=\qty\Big{G_0(E+\epsilon/2)+\sum_{n=1}^\infty B^n \frac{1}{4\pi in}\int_0^\beta d\beta\;\beta \int_{-i\infty}^{i\infty}ds  e^{(E-s-1/2+\epsilon/2)\beta}(I(s,\epsilon)I(s,0))^n}+\qty\Big{(\epsilon\rightarrow-\epsilon)}\nonumber \\
  &=\qty\Big{G_0(E+\epsilon/2)+\sum_{n=1}^\infty B^n \frac{1}{4\pi in}\int_0^\beta d\beta\;\pdv{E} \int_{-i\infty}^{i\infty}ds e^{(E-s-1/2+\epsilon/2)\beta}(I(s,\epsilon)I(s,0))^n}+\qty\Big{(\epsilon\rightarrow-\epsilon)}\nonumber \\
  &=\qty\Big{G_0(E+\epsilon/2)+\sum_{n=1}^\infty B^n \frac{1}{4\pi in}\pdv{E}\int_{-i\infty}^{i\infty}ds \frac{1}{s-E+1/2-\epsilon/2}(I(s,\epsilon)I(s,0))^n}+\qty\Big{(\epsilon\rightarrow-\epsilon)}\nonumber \\
  &=\qty\Big{G_0(E+\epsilon/2)+\frac{1}{2}\sum_{n=1}^\infty B^n \frac{1}{n}\pdv{E}(I(s=E-1/2+\epsilon/2,\epsilon)I(s=E-1/2+\epsilon/2,0))^n}+\qty\Big{(\epsilon\rightarrow-\epsilon)},
\end{align}
where $G_0(E)=\pdv{E}\log\Gamma(1/2-E)$ is the resolvent of harmonic oscillator. (See Appendix \ref{Fredholm_harmonic})\\
Using $-\log(1+x)=\sum_{n=1}^{\infty}\frac{(-x)^n}{n}$ gives
\begin{align}
  \label{resolvent}
  G(E)=&-\pdv{E}\qty\Big{-\log\Gamma\qty(\frac{1}{2}-E-\frac{\epsilon}{2})+\frac{1}{2}\log(1-BI(s=E-1/2+\epsilon/2,\epsilon)I(s=E-1/2+\epsilon/2,0))}\nonumber \\
  &-\pdv{E}\qty\Big{(\epsilon\rightarrow-\epsilon)}
\end{align}
Using $G(E)=-\pdv{E}\log D$, we obtain the Fredholm determinant as follows
\begin{align}
  D(E)=\frac{1}{\Gamma\qty(\frac{1}{2}-E-\frac{\epsilon}{2})\Gamma\qty(\frac{1}{2}-E+\frac{\epsilon}{2})} \sqrt{1-BI(s_{+\epsilon},\epsilon)I(s_{+\epsilon},0)}\sqrt{1-BI(s_{-\epsilon},-\epsilon)I(s_{-\epsilon},0)},\label{DDDD}
\end{align}
where $s_{\pm\epsilon}=E-1/2\pm\epsilon/2$. Substituting $I(s,\epsilon)=e^{\pm i\pi (\epsilon-s)}\qty(\frac{\hbar}{2})^{\epsilon-s}\Gamma(\epsilon-s)$ into (\ref{DDDD}), $D(E)$ becomes
\begin{align}
  D(E)=\frac{1}{\Gamma\qty(\frac{1}{2}-E-\frac{\epsilon}{2})\Gamma\qty(\frac{1}{2}-E+\frac{\epsilon}{2})}\qty(1-B e^{\pm i\pi (1-2E)}\qty(\frac{\hbar}{2})^{1-2E}\Gamma\qty(\frac{1}{2}-E-\frac{\epsilon}{2})\Gamma\qty(\frac{1}{2}-E+\frac{\epsilon}{2}))
\end{align}

Therefore, $D(E)=0$ gives
\begin{align}
  \label{FredD}
  \frac{1}{\Gamma\qty(\frac{1}{2}-E-\frac{\epsilon}{2})\Gamma\qty(\frac{1}{2}-E+\frac{\epsilon}{2})}-Be^{\pm i\pi\qty(1-2E)}\qty(\frac{\hbar}{2})^{\qty(1-2E)}=0,
\end{align}
where $B=\frac{e^{-\frac{1}{3\hbar}}}{2\pi}\frac{2}{\hbar}$.
If we only consider the finite number of bions, the partition function (\ref{Form,Z}) is still singular like $\sim 1/\epsilon^n$. However, using the analytic continuation ($-\log(1+x)=\sum_{n=1}^{\infty}\frac{(-x)^n}{n}$) give analytic function around $\epsilon=0$. Therefore the summation of all classical periodic solutions is necessary to solve the $1/\epsilon$ problem.

\section{Calculating the ground state energy}
\subsection{For $\epsilon=0$, symmetric double well}
\begin{align}
  \frac{1}{\Gamma\qty(\frac{1}{2}-E)\Gamma\qty(\frac{1}{2}-E)}-Be^{\pm i\pi\qty(1-2E)}\qty(\frac{\hbar}{2})^{\qty(1-2E)}=0
  \label{ep0}
\end{align}
We can write the ground state energy as $E=\frac{1}{2}+x$, where $x$ is exponential small factor. Then (\ref{ep0}) becomes
\begin{align}
  \frac{1}{\Gamma(-x)}&=\sqrt{B}e^{\mp\pi ix}\qty(\frac{\hbar}{2})^{-x}\\
  \frac{1}{\Gamma(-x)}&=-\sqrt{B}e^{\mp\pi ix}\qty(\frac{\hbar}{2})^{-x}
\end{align}
The $\pm$ corresponds to the parity. Using the reflection formula: $\Gamma(x)\Gamma(1-x)=\frac{\pi}{\sin \pi x}$, 
\begin{align}
  \frac{\sin\pi x}{\pi}&=\sqrt{B}e^{(\mp\pi i-\log \frac{\hbar}{2})x}\frac{1}{\Gamma(1+x)}\\
  \frac{\sin\pi x}{\pi}&=-\sqrt{B}e^{(\mp\pi i-\log \frac{\hbar}{2})x}\frac{1}{\Gamma(1+x)}
\end{align}
 Finally, the nonperturbative contribution is
\begin{align}
  x&=\sqrt{B}\qty(1+\qty(\mp\pi i-\log\qty(\frac{\hbar}{2})+\gamma)x+O(x^2))\\
  &=\sqrt{B}+\qty(\mp\pi i-\log\qty(\frac{\hbar}{2})+\gamma)B+O(B^{3/2})\\
  \text{and}\\
  x&=-\sqrt{B}-\qty(\mp\pi i-\log\qty(\frac{\hbar}{2})+\gamma)B+O(B^{3/2})
  \label{ambiguityE0}
\end{align}
Therefore, the result gives indeed energy splitting by one instanton: $\sqrt{B}$, and the imaginary ambiguity is proportional to $B$ (bion) for the symmetric double well.
The ambiguity of ground state energy(\ref{ambiguityE0}) is exactly cancelled by Borel ambiguity from the perturbative expansion around the vacuum. (set $\epsilon=0$ in (\ref{SUSYBorelambiguity}))
\subsection{For $\epsilon=1$, SUSY case}
The Fredholm determinant (\ref{FredD}) is valid for any $\epsilon$ and we can show the correct nonperturbative effect for $\epsilon=1$. 
\begin{align}
  \frac{1}{\Gamma\qty(-E)\Gamma\qty(1-E)}-Be^{\pm i\pi\qty(1-2E)}\qty(\frac{\hbar}{2})^{\qty(1-2E)}=0
\end{align}
We know the pertubative part of the ground state energy is $0$. Setting $E=0+x$ gives
\begin{align}
  \frac{1}{\Gamma\qty(-x)\Gamma\qty(1-x)}-Be^{\pm i\pi\qty(1-2x)}\qty(\frac{\hbar}{2})^{\qty(1-2x)}=0
\end{align}
This is rewritten as
\begin{align}
  \frac{\sin\pi x}{\pi}=B\frac{\hbar}{2}e^{\mp 2\pi i x}\qty(\frac{\hbar}{2})^{-2x}
\end{align}
This is solved as 
\begin{align}
  x&=B\frac{\hbar}{2}+O(x)\\
   &=\frac{e^{-1/3\hbar}}{2\pi}+O(B^2)
\end{align}
The leading nonpertubative contribution is due to a bion and coincides to (\ref{CBE02}). 

\section{Discussion and summary}
In this work, we showed the $1/\epsilon$ problem in the tilted double well potential is solved by including multi-complex bion and multi-bounce contributions. The result suggests the relationship between exact WKB and path integral formalism.

\subsection{Conjecture about the relation between exact WKB method}
We conjecture the form of the Fredholm determinant is equivalent to the quantization condition derived from exact WKB method.

At leading order of the exact WKB calculation, we can evaluate the quantization condition from the connection formula of symmetric double well
\begin{align}
  \sim (1-e^{2\pi ix})^2\qty(1-\frac{\sqrt{B}}{1-e^{2\pi i x}})\qty(1+\frac{\sqrt{B}}{1-e^{2\pi i x}})=0\;\;\;\qty(x=\frac{1}{2}-E)
\end{align}
$(1-e^{2\pi ix})$ comes from Voros coefficient related to harmonic oscillator(2 nondegenerate Stokes curves with a simple turning point) and the other part $(1-\frac{\sqrt{B}}{1-e^{2\pi i x}})(1+\frac{\sqrt{B}}{1-e^{2\pi i x}})$ comes from an infinite number of Borel singularities at $z=2 \pi n E$ ($n\in \mathbb{Z}$) in the Borel plane, which are called fixed singularities in exact WKB literatures. The factor $B$ is a nonperturbative term ($\sim e^{-\frac{A}{g}}$) from the other Voros coefficient. 

when we compare the distribution of zeros, we can assume $\frac{1}{\Gamma(x)}\sim 1-e^{2\pi i x}$ and $1-\Gamma(x)\sim 1-\frac{1}{1-e^{2\pi ix}}$. If this identification is verified, we can assume the Fredholm determinant via path integral is equivalent to the quantization condition from exact WKB.
Furthermore, it suggests the reason why $1/\epsilon$ problem is solved considering multi-bion configuration. To obtain the correct quantization condition, we have to take into account the infinite number of Borel singularities, which are corresponded to multi-bion.

Also this correspondence suggests a method to calculate the intersection number of Lefschetz thimble with exact WKB:
\begin{align}
  Z&=\tr e^{-\beta\hat{H}}\\
  &=\int \mathcal{D}x\;e^{-\frac{S[x]}{\hbar}}\\
  &=\mathcal{S}[e^{-\frac{S[x_0]}{\hbar}}\sum a_n\hbar^n]+\mathcal{S}[e^{-\frac{S[x_1]}{\hbar}}\sum b_n\hbar^n]+...\\
  &=\sum_\sigma n_\sigma \; \int_{\mathcal{J}_\sigma} \mathcal{D}x\;e^{-\frac{S[x]}{\hbar}}=\sum_\sigma n_\sigma\;Z_\sigma(\beta)
\end{align}
$\mathcal{S}$ denotes the Borel summation of each series and $x_\sigma$ is a saddle point. $\mathcal{J}_\sigma$ and $n_\sigma$ are called Lefschetz thimble and intersection number in Picard Lefschetz theory.
The Laplace transform of $Z(\beta)$ gives the trace of resolvent $G(E)$ but this is linear transform, therefore we can write 
\begin{align}
  \tr\frac{1}{H-E}=G(E)&=\int_0^\infty Z(\beta)e^{\beta E}\dd{\beta}\\
  &=\sum_\sigma n_\sigma \int_0^\infty Z_\sigma(\beta)e^{\beta E}\dd{\beta}\\
  &=\sum_\sigma n_\sigma G_\sigma(E)
\end{align}
The trace of resolvent $G(E)$ can be expressed as $-\pdv{E}\log D=G(E)$, it means
\begin{align}
  D(E)=\prod_\sigma D_\sigma^{n_\sigma}(E)
\end{align}
Therefore if this $D(E)$\footnote{The definition of Fredholm determinant (or resolvent) needs a regularization, e.g. $G_{reg.}\equiv G(E)-G(0)$ or $D_{reg.}\equiv \frac{D(E)}{D(0)}$ or zeta function regularization for $D(E)$} is equivalent to the quantization condition derived from exact WKB, we can calculate the intersection number $n_\sigma$ with this method.

\section*{Acknowledgements}
The author appriciates Tatsuhiro Misumi, Norisuke Sakai, Toshiaki Fujimori, Mithat \"{U}nsal, Alireza Behtash and Yuya Tanizaki for useful discussion and comments. The author specially thanks T. Misumi for discussion of the $1/\epsilon$ problem for a long time. I acknowledge financial support from KMI, Nagoya University.
\begin{appendices}
  \label{DIG}
  \section{About the leading contribution of symmetric double well potential in dilute gas approximation}
  The partition function must satisfy with periodic boundary condition. Therefore
  \begin{align}
    Z= 2(Z_0+Z_{bion}+Z_{2-bion}+...)
  \end{align}
  The factor 2 comes from the two vacua. The bion contributions are nonperturbative effects. So we often consider $Z_0>Z_{bion(s)}$ in weak coupling limit.
  Under this assumption we can calculate the ground state energy like this:
\begin{align}
  E_0=-\frac{1}{\beta}\log Z&=-\frac{1}{\beta}\log(Z_0+Z_{bion}+...)\\
  &=-\frac{1}{\beta}\log\qty(Z_0\qty(1+\frac{Z_{bion}}{Z_0}+...))\\
  &=-\frac{1}{\beta}\log Z_0-\frac{1}{\beta}\log\qty(1+\frac{Z_{bion}}{Z_0}+...)\\
  &=-\frac{1}{\beta}\log Z_0-\frac{1}{\beta}\qty(\frac{Z_{bion}}{Z_0})+...\;\;\;\;\;\qty(\text{when }\frac{Z_{bion(s)}}{Z_0}<1)
  \label{formula_E0}
\end{align}
Therefore the leading nonperturbative contribution is coming from one bion in this calculation. In fact, this method is quite ordinal in many literatures. Although, it gives a wrong answer in this case. As well known, the leading nonperturbative contribution is one instanton but not one bion. This contradiction is because of the incorrect assumption: $\frac{Z_{bion(s)}}{Z_0}<1$.

As I showed in (\ref{DIGbion}), The partition function of this system is $\frac{Z}{Z_0}=2\sum_{n=0}^{\infty}\frac{B^n\beta^{2n}}{(2n)!}=2\cosh(\sqrt{B}\beta)$. This expression contatins the multi-bion. It gives
\begin{align}
  E_0&=-\frac{1}{\beta}\log Z\\
  &=-\frac{1}{\beta}\log Z_0-\frac{1}{\beta}\log (e^{\sqrt{B}\beta}+e^{-\sqrt{B}\beta})-\frac{1}{\beta}\log 2\\
  &=-\frac{1}{\beta}\log Z_0-\frac{1}{\beta}\log (e^{\sqrt{B}\beta}(1+e^{-2\sqrt{B}\beta}))-\frac{1}{\beta}\log 2\\
  &=-\frac{1}{\beta}\log Z_0-\sqrt{B}\;\;\;(\text{in }\beta\rightarrow\infty)
\end{align}
Therefore the leading contribution is one instanton. This puzzle comes from the $\beta$, which comes from the translation symmetry of classical solution. This $\beta$ is infinity to calculate the ground state energy, so the assumption $\frac{Z_{bion(s)}}{Z_0}<1$ is invalid generally.
\section{Resolvent of harmonic oscillator}
\label{Fredholm_harmonic}
Consider harmonic oscillator. i.e. the eigenvalues are $\frac{1}{2}+n\;\;\;(n=0,1,2...)$
Then the Fredholm determinant is
\begin{align}
  D(E)=\det(H-E)=\prod_{n=0}^\infty\qty(n+\frac{1}{2}-E)
\end{align}
This infinite product is ill-defined though, we can define this quantity with 
zeta function regularization(zeta regulated product).

In this case, the spectral zeta function is
\begin{align}
  \zeta\qty(s)=\sum_{n=0}^\infty \frac{1}{\qty(n+\frac{1}{2}-E)^s}
\end{align}
This is Hurwitz zeta function. Also we get
\begin{align}
  \zeta(s,a)=\sum_{n=0}^\infty \frac{1}{(n+a)^s}\\
  \eval{\pdv{s}\zeta(s,a)}_{s=0}=\log\Gamma(a)-\frac{1}{2}\log(2\pi)
\end{align}
Therefore the determinant is evaluated in terms of zeta regularized product,
\begin{align}
  D(E)=\det(H-E)&=e^{-\zeta'(0,\frac{1}{2}-E)}\\
  &=\frac{\sqrt{2\pi}}{\Gamma\qty(\frac{1}{2}-E)}
\end{align}
The trace of the resolvent is
\begin{align}
  G(E)=-\pdv{E}\log D=\pdv{E}\log\Gamma\qty(\frac{1}{2}-E)
\end{align}
(The constant term $\sqrt{2\pi}$ is dropped by the derivative)\\
The zeros of $D(E)$ are the poles of $\Gamma\qty(\frac{1}{2}-E)$. i.e.the eigenvalue of harmonic oscillator: $\frac{1}{2}+n\;\;\;(n=0,1,2...)$

\end{appendices}

\bibliographystyle{ieeetr}
\bibliography{epsilon_problem_v2}

\end{document}